%Paper: hep-th/9312069
%From: segui@cc.unizar.es
%Date: Thu,  9 DEC 93 13:50 GMT

\baselineskip=18pt
\magnification=1200
\centerline{\bf A MODIFIED SCHWINGER'S FORMULA}
\bigskip
\centerline{\bf FOR THE CASIMIR EFFECT}
\vskip 0.5cm
\centerline{\bf M.V. Cougo-Pinto, C.Farina}
\centerline{\it Instituto de F\'\i sica-Universidade Federal do Rio de Janeiro}
\centerline{\it Rio de Janeiro, R.J., CEP - 21.845-970 - Brasil}
\bigskip
\centerline{\bf Antonio J. Segu\'\i-Santonja}
\centerline{\it Departamento de F\'\i sica Te\'orica}
\centerline{\it Universidad de Zaragoza - Zaragoza 50009 - Spain}
\centerline{ (A short version will appear in the Letters in Math. Phys.)}
\vskip 0.5cm
\centerline{\bf Abstract}
\bigskip

After briefly reviewing how the (proper-time) Schwinger's formula
works for computing the Casimir energy in the case of \lq\lq scalar
electrodynamics" where the boundary conditions are dictated by two
perfectly conducting parallel plates with separation $a$ in the
$Z$-axis, we propose a slightly modification in the previous approach
based on an analytical continuation method. As we will see, for the case
at hand our formula does not need the use of Poisson summation to get
a (renormalized) finite result.
\vfill\eject

In a recent letter Julian Schwinger$^{(1)}$ has taught us how to
obtain the Casimir energy from the Schwinger formula for the
effective action in the proper-time representation$^{(2)}$:
$$W^{(1)}(s_0)=-{i\over2}\int_{s_0}^\infty{ds\over s}Tr e^{-isH}
\; +\; const.,\eqno(1)$$
where the exponential of $iW^{(1)}$ is defined as the vacuum
persistency probability amplitude $\langle 0_+\vert 0_-\rangle$ at
the one-loop level, $Tr$ means the total trace, $H$ is the
corresponding proper-time Hamiltonian and $s_0$ is a regularization
cutoff that must be sent to zero only at the end of the
calculations, after an appropriate subtraction of divergent terms is
made; the additional constant is used to subtract divergent terms,
thereby establishing the physical normalization for the situation.

Let us briefly show how equation (1) can be used to yield the correct
Casimir energy in a simple example. Consider the situation of two
perfectly conducting parallel plates with separation $a$ along the
$Z$-axis. In this case we have
$$H=\left[ P_x^2+P_y^2-\omega^2\right]+P_z^2,\eqno(2)$$
where $\vec P=-i\nabla$ and $\omega=+i\partial_t$. Observing that,
while the operators in the bracket on the r.h.s. of (2) have
continuous spectra, the operator $P_z^2$ has a discretized one, given
by \break $p_z^2=\left({n\pi\over a}\right)^2$ ; $n=1,2,...$, a
straightforward calculation shows that
$$Tr\, e^{-isH}=AT{1\over \sqrt{i(4\pi s)^3}}\, \sum_{n=1}^\infty \,
e^{-is\left({n\pi\over a}\right)^2}\; ,\eqno(3)$$
where $A$ is the (finite, although as big as we want) area of each
plate and $T$ the finite duration of the measurement.

After identifying the vacuum energy shift ${\cal E}$ with $-{W^{(1)}\over
T}$, we obtain from (1) and (3) that
$${{\cal E}(s_0;a)\over A}={\sqrt{i}\over 16\pi^{3/2}}\,\int_{s_0}^\infty \,
\; {ds\over s^3}
\sum_{n=1}^\infty \, e^{-in^2 s(\pi/a)^2}.\eqno(4)$$

It is obvious that the limit $s_0\rightarrow 0$ gives a divergent
result so that we need to separate the infinities from the
relevant finite part of equation (4). With this goal, we shall use
the Poisson summation formula,
$$\sum_{n=-\infty}^\infty \, e^{-n^2\pi\tau}=\sqrt{{1\over \tau}}\,
\sum_{n=-\infty}^\infty \, e^{-n^2\pi (1/\tau)}\, .\eqno(5)$$
It follows from (5) that
$$\sum_{n=1}^\infty \, e^{-in^2 s(\pi/a)^2}= {a\over2}{1\over
\sqrt{i\pi s}}-{1\over2}+{a\over \sqrt{i\pi s}}\, \sum_{n=1}^\infty
\, e^{i(na)^2/s}\, .\eqno(6)$$

Inserting (6) into (4), we have
$$\eqalignno{{{\cal E}(s_0;a)\over A}=&{1\over A}(Aa){1\over
32\pi^2}\int_{s_0}^\infty {ds\over s^3}-{1\over A}A{\sqrt{i}\over
32\pi^{3/2}} \int_{s_0}^\infty {ds\over s^{5/2}}+\cr
&+{a\over 16\pi^2}\int_{s_0}^\infty {ds\over s^3}\, \sum_{n=1}^\infty
\; e^{i(na)^2/s}\; .&(7)\cr}$$

The first two terms on the r.h.s. of (7) are divergent when
the limit $s_0\rightarrow 0$ is taken, and hence they must be
discarded. The physical reason for that is the following: the first
one is proportional to the spatial volume $Aa$ between the two
plates, giving a uniform (infinite) contribution for the Casimir
energy which is independent of the positions of the plates. Since
this term violates the correct normalization of ${\cal E}$ (${\cal
E}=0$ for $a\rightarrow\infty$), we must absorb it in the
normalization of $\langle 0_+\vert 0_-\rangle$. The second one does
not contain $a$, so that it corresponds to the (infinite) energy
required to create both plates; since we are only interested in
energy shifts caused by changing the relative configuration of the
plates we must also disregard this contribution.

Therefore, we are left with
$$\eqalignno{{{\cal E}(a)\over A}&=\lim_{s_0\rightarrow 0}\; {a\over
16\pi^2} \, \int_{s_0}^\infty {ds\over s^3}\, \sum_{n=1}^\infty \,
e^{i(na)^2/s}\cr
&=-{1\over 16\pi^2}\, {1\over a^3}\, \sum_{n=1}^\infty\int_0^\infty
\, d\sigma\, \sigma \, e^{-n^2\sigma}\; , &(8)\cr}$$
where we made the convenient change of variables $\sigma={a^2\over is}$.

Appealing to the well known formulas
$$\int_0^\infty \, dz\, z^{\nu-1} \, e^{-Az}=\Gamma(\nu)A^{-\nu}\;\;\;
\;\;\; (Re(\nu)>0),\eqno(9a)$$
$$\zeta_R(s)=\sum_{n=1}^\infty\, n^{-s}\;\;\; \;\;\; (Re(s)>1),\eqno(9b)$$
where $\zeta_R(s)$ is the Riemann zeta function, we get
$$\eqalignno{{{\cal E}(a)\over A}&=-{1\over 16\pi^2}\, {1\over
a^3}\zeta_R(4)\cr
&=-{\pi^2\over 1440}\, {1\over a^3}\, ,&(10)\cr}$$
which, apart from the factor 2 that takes into account the two photon
polarizations (this simple multiplication by 2 does not work when
other geometries are involved), is the correct result found by other
methods$^{(3)}$.

Now, we shall show that the Schwinger's method for a massless scalar
theory, with a slight modification in the regularization process,
leads to the finite result in an exceedingly short and simple way,
without requiring any subtraction of infinities. The appeareance of
any divergent term during the calculation is bypassed by means of
analytical continuation of the functions involved.

The basic idea of our approach is the following: since the origin of
the divergent result in (1) when taking $s_0\rightarrow 0$ was the
presence of negative powers of $s$ in the integrand, we can modify
equation (1) by including as many positive powers of $s$ as necessary
(in fact we shall include the factor $s^\nu$) so that we will be able
to release the lower limit of the integral to zero. Therefore, our
starting point is given by
$$W^{(1)}(\nu)=-{1\over2}\int_0^\infty \, {ds\over s} s^\nu \,
Tr\, e^{-isH},\eqno(11)$$
where $\nu$ must be sufficiently large to regularize the integral.
After the integral is evaluated, we make an analytical extension of
$W^{(1)}(\nu)$ to the whole complex plane (of $\nu$) and then take
the limit $\nu\rightarrow 0$. As we will see, for the case at hand,
this procedure will give directly the desired (finite) result.

Inserting (3) into (11), we have (after identifying ${\cal
E}=-{W^{(1)}\over T}$)
$$\eqalignno{{{\cal E}(\nu;a)\over A}&={\sqrt{i}\over 16\pi^{3/2}}\,
\sum_{n=1}^\infty\int_0^\infty \, ds\; s^{\nu-3/2-1}\;
e^{-s(i\pi^2 n^2/a^2)}\cr
&=-{\pi^2\over 16}{1\over \sqrt{\pi}}\Gamma(\nu-3/2)\zeta_R(2\nu-3)
{1\over a^3}\; ,&(12)\cr}$$
where we used equations (9a) and (9b). To finish the calculation we
take the limit $\nu\rightarrow 0$ and substitute the value of
$\Gamma(-3/2)\zeta_R(-3)$. The result is
$${{\cal E}(a)\over A}=-{\pi^2\over 1440}{1\over a^3},\eqno(13)$$
in complete agreement with (10).

Let us finish this paper with two final remarks. First, that this
last method works as well (with no need of subtracting infinities) in
arbitrary $d+1$-dimensional space-time, leading us quickly to the
well known result$^{(3)}$:
$${{\cal E}\over A}=-{1\over2}(\sqrt{\pi}/2)^d\Gamma(-{d\over2})
\zeta_R(-d){1\over a^d}.\eqno(13)$$
However, attention must be paid to the cases of even spatial
dimensions (even $d$), where poles of the gamma function are
cancelled by the zeros of the Riemann zeta function. Second, if our
regularization, equation (11),
method is applied to other theories, as QED with constant external
fields, or even in the computation of the Casimir energy for a
massive scalar field,
subtraction of infinities may also be required
 before the analytical continuation to $\nu=0$ is made. In
a more detailed analysis, to be reported elsewhere, we will further
explore these remarks.

We would like to thank to M. Asorey for a critical reading of the
manuscript. This work was partially supported by MEC-CAICYT and
Conselho Nacional de Desenvolvimento Cientifico e Tecnologico (CNPq).
One of us (A.J.S.S.) was also supported by DGICYT(Spain), grant PB90-0916.

\vfill\eject
\centerline{\bf References}

\item{1.} Schwinger, J., {\it Lett. Math. Phys.} {\bf 24}, 59 (1992).
\item{2.} Schwinger, J., {\it Phys. Rev.} {\bf 82}, 664 (1951).
\item{3.} Plunien, G., Muller, B. and Greiner, W., {\it Phys. Rep.}
{\bf 134}, 87 (1986).

\bye